\documentclass[11pt]{article}
\usepackage{times}

\input{tcilatex}
\begin{document}

\title{Parallel Programming with Matrix Distributed Processing}
\author{Massimo Di Pierro \\
{{\footnotesize School of \textbf{C}omputer Science, \textbf{T}%
elecommunications and \textbf{I}nformation Systems}}\\
{\footnotesize DePaul University, 243 S. Wabash Av., Chicago, IL 60604, USA}}
\maketitle

\begin{abstract}
Matrix Distributed Processing (MDP)\ is a C++ library for fast development
of efficient parallel algorithms. \texttt{MDP} enables programmers to focus
on algorithms, while parallelization is dealt with automatically and
transparently. Here we present a brief overview of \texttt{MDP} and examples
of applications in Computer Science (Cellular Automata), Engineering (PDE
Solver) and Physics (Ising Model).
\end{abstract}

\section{Introduction}

Matrix Distributed Processing (\texttt{MDP}) \cite{mdp1}\cite{mdp2} is a
collection of classes and functions written in C++ for fast development of
parallel algorithms such as solvers for partial differential equations,
mesh-like algorithms, and various types of graph-based problems. These
algorithms find frequent application in many sectors of physics,
engineering, electronics and computational finance.

\texttt{MDP} includes:

\begin{itemize}
\item  a natural syntax for the algorithms that is transparent to the
underlying parallelization;

\item  parallelization based optimization algorithms implemented in MPI;

\item  functions for linear algebra computations with support of a
Maple-like syntax;

\item  statistical functions;

\item  fitting functions;

\item  a Parallel SIMulator (PSIM) which allows \texttt{MDP} algorithms to
run on single processor machines without MPI (uses fork for creating the
processes and socket pairs for communication).
\end{itemize}

\texttt{MDP} was originally developed for and still constitutes the core of
FermiQCD\cite{fermiqcd2002}\cite{fermiqcd2003}, a library for Lattice QCD
computations. Lattice QCD is a Monte Carlo based numerical approach to the
physics of composite particles made of quarks (e.g., protons and neutrons)
and it is considered one of the most computationally intensive projects of
modern physics.

FermiQCD, was developed by the author at the University of Southampton (UK)
and Fermilab (Department of Energy) and it is currently used by other
physics departments around the world\cite{app1}\cite{app2}\cite{app3}\cite
{app4}\cite{app5}.

While Lattice QCD is currently one of the main applications of \texttt{MDP},
its range of applicability is not limited to it as we show in the following
examples.

\section{Example: Parallel Game of Life}

As a first example we show here how to write, in a few lines, a parallel
program to find stable configurations for the Game Of Life. The game works
as follows:

\begin{itemize}
\item  It is defined on a board ($N\times N$ cells) with periodic boundary
conditions.

\item  Each cell can be alive (1) or dead (0).

\item  The game consists of a series of iterations and, at each iteration,
if a cell is dead and the number of beighbor cells who are alive is greater
or equal 3, the cell dies; if a cell is alive and the number of beighbor
cells who are alive is 2 or 3, the cell remains alive; otherwise the cell
will be dead.

\item  Given a random starting configuration for the board, we want to
determine its evolution and whether it reaches a stable configuration.
\end{itemize}

Here is the entire parallel code:
\begin{verbatim}
00  #include "mdp.h"         
01
02  const int alive=1, dead=0;         // some constants
03
04  int rulesofgame(int a00, int a01, int a02,
05                  int a10, int a11, int a12,
06                  int a20, int a21, int a22) {
07    int sum=a00+a01+a02+a10+a12+a20+a21+a22;
08    if(a11==dead && sum==3) return alive;
09    else if(a11==alive && (sum==2 || sum==3)) return alive;
10    return read;
11  }                                  // rules of the game         
12
13  int main(int argc, char **argv) {
14    mdp.open_wormholes(argc,argv);   // open communication
15    int L[]={10,10};                 // declare board size
16    mdp_lattice board(2,L);         // declare board
17    mdp_field<int> C(board);         // declare cells
18    mdp_field<int> newC(board);      // declare new set of cells
19    mdp_site x(board);             
20
21    forallsites(x) C(x)=(board.random(x).plain()>0.5)?alive:dead;
22
23    while(1) {                       // game iteration
24      C.update();                    // communicate!
25      forallsites(x)                 // parallel loop
26        newC(x)=rulseofgame(C((x-1)-0),C(x-1),C((x-1)+0),
27                            C(x-0),    C(x),  C(x+0),
28                            C((x+1)-0),C(x+1),C((x+1)+0));  
29      int diff=0;
30      forallsites(x) {
31        diff+=abs(C(x)-newC(x));     // count changed cells
32        C(x)=newC(x);                // store new cells
33      }
34      if(diff==0) break;             // exit if cells didn't change
35    }
36    C.save("cells.dat")              // save cells
37    mdp.close_wormholes();
38    return 0;
39  }
\end{verbatim}

\begin{itemize}
\item  Line 00 includes the basic library.

\item  Lines 04-11 define the rule of the game. The values of a cell (%
\texttt{a11}) and its neighbors are passed as a $3\times 3$ table (\texttt{%
a00-a22}). The function returns the new value for \texttt{a11}.

\item  Line 14 opens the parallel communication channels, line 37 closes
them.

\item  Line 15 declares a 2D array containing the board size, \texttt{L}.

\item  Line 16 declares a lattice object (board) that represents the set of
board sites and its topology. By default a mesh topology with toroid
boundary conditions. Two arguments are passed to the constructor: the
dimension, and size of the board. For each board site, it is possible to
specify on which parallel process it is going to be allocated as well as
alternative topologies.

\item  Line 17 declares a field of cells (\texttt{C}) on the board (the
cells that live on it).

\item  Line 18 declares an auxiliary field that will be necessary for the
computation.

\item  Line 19 declares a \texttt{x} a variable that represents a generic
site of the board and will be used for looping over the sites and therefore
the cells that live on them.

\item  Line 21 sets the initial configuration of the board by looping over
all sites \texttt{x} and setting the corresponding cell \texttt{C(x)} to a
random \texttt{alive} or \texttt{dead}. Each process loops only over the
sites that are allocated locally by the process. Notice how MDP provides a
local random number generator for each site of the board, \texttt{%
board.random(x)}, which is vital in order to be able to reproduce results of
stochastic algorithms.

\item  Lines 23-35 loop over the iterations of the game.

\item  Line 24 is performs a critical operation; it informs MDP that the
value of the field \texttt{C} has changed and parallel communication is
needed to synchronize those buffers that contain copies of non-local sites (%
\emph{synchronization}).

\item  Lines 25-28 apply the rules of the game at each site. The new values
of the states for the cells are stored in the auxiliary field \texttt{newC}.
Each process loops over the local sites only. For each local site (\texttt{x}%
), \texttt{x+0} represents the neighbor site when coordinate 0 is
incremented by one, \texttt{x-0} represents the neighbor site when
coordinate 0 is decremented by one, \texttt{x+1} represents the neighbor
site when coordinate 1 is incremented by one, etc.\footnote{%
This notation may appear bizzarre but it is nothing more than a sum ($%
\overrightarrow{x}+\overrightarrow{i}$) or difference ($\overrightarrow{x}-%
\overrightarrow{i}$) of vectors where $\overrightarrow{x}$ is a vector that
represents a site on the board and the integer $\overrightarrow{i}$
represents a unit verson in direction $i.$}.

\item  Lines 29-33 perform two operations in a single parallel loop: the new
states for the cells (\texttt{newC}) are copied back into \texttt{C}; the
number of cells that have changed state are counted and the number is stored
in \texttt{diff}.

\item  Line 34 terminates the iteration if no cell has changed its state.

\item  Line 36 saves the board configuration in a file \texttt{cells.dat}.
\end{itemize}

Most of the parallel work is done by the constructor (\texttt{mdp\_lattice})
of the \texttt{board} which, from the board topology, determines how to
partition it and determines the optimal communication patterns; by the
constructor (\texttt{mdp\_field}) of the field of cells (\texttt{C}) which
allocates the local cells and the buffers to store copies of non-local
cells; and by the method \texttt{update} which performs communication to
copy remote cell values into the local buffers.

We'll show in the next few examples that is equally easy to implement fields
of any class of objects. In the next example we'll consider a field of
matrices.

\section{Example: Parallel PDE Solver}

Consider here, as a different example that presents many similarities with
the one above, the following Laplace equation: 
\begin{equation}
\nabla ^2\varphi (x)=f(x)  \label{eq1}
\end{equation}
where $\varphi (x)$ is a field of $2\times 2$ complex matrices defined on a
3D space (\texttt{space}), $x=(x_0,x_1,x_2)$ limited by $0\leq x_i<L_i$, and 
\begin{eqnarray}
L &=&\{10,10,10\} \\
f(x) &=&A\sin (2\pi x_1/L_1)  \nonumber \\
A &=&\left( 
\begin{array}{ll}
1 & i \\ 
3 & 1
\end{array}
\right)  \nonumber
\end{eqnarray}
The initial conditions are $\varphi _{initial}(x)=0.$ We will also assume
that $x_i+L_i=x_i$ (torus topology).

\textbf{Solution:} In order to solve eq.~(\ref{eq1}) we first discretize the
Laplacian ($\nabla ^2=\partial _0^2+\partial _1^2+\partial _2^2$) and
rewrite it as 
\begin{equation}
\sum_{\mu =0,1,2}\left[ \varphi (x+\widehat{\mu })-2\varphi (x)+\varphi (x-%
\widehat{\mu })\right] =f(x)
\end{equation}

where $\widehat{\mu }$ is a unit vector in the discretized space in
direction $\mu $. Hence we solve for $\varphi (x)$ and obtain the following
recurrence relation 
\begin{equation}
\varphi (x)=\frac{\sum_{\mu =0,1,2}\left[ \varphi (x+\widehat{\mu })+\varphi
(x-\widehat{\mu })\right] -f(x)}6  \label{eq3}
\end{equation}

The following is a typical \texttt{MDP} program that solves eq.~(\ref{eq1})
by recursively iterating eq.~(\ref{eq3}). Notice how the program is parallel
but there are no explicit calls to communication functions:
\begin{verbatim}
00  #include "mdp.h"
01
02  void main(int argc, char** argv) {
03     mdp.open_wormholes(argc,argv);    // open communications
04     int L[]={10,10,10};               // declare volume
05     mdp_lattice      space(3,L);      // declare lattice
06     mdp_site         x(space);        // declare site variable
07     mdp_matrix_field phi(space,2,2);  // declare field of 2x2
08     mdp_matrix       A(2,2);          // declare matrix A
09     A(0,0)=1;  A(0,1)=I;
10     A(1,0)=3;  A(1,1)=1;
11     forallsites(x)                    // loop (in parallel)
12        phi(x)=0;                      // initialize the field
13     phi.update();                     // communicate!
14
15     for(int i=0; i<1000; i++) {       // iterate 1000 times
16        forallsites(x)                 // loop (in parallel)
17           phi(x)=(phi(x+0)+phi(x-0)+
18                   phi(x+1)+phi(x-1)+
19                   phi(x+2)+phi(x-2)-
20                   A*sin(2.0*Pi*x(1)/L[1]))/6;  // the equation
21        phi.update();                  // communicate!
22     }
23     phi.save("field_phi.mdp");        // save field
24     mdp.close_wormholes();            // close communications
25  }
\end{verbatim}

\textbf{Notes:}

\begin{itemize}
\item[$\bullet$]  Line 04 declares the size of the box used to approximate
the space $L=\{L_0,L_1,L_2\}$.

\item[$\bullet$]  Line 05 declares a 3-dimensional lattice, called \texttt{%
space}, on the box $L$. \texttt{MDP} supports up to 10-dimensional lattices.
By default, a lattice object is a mesh with torus topology.

\item[$\bullet$]  Line 06 declares a site variable site \texttt{x} that will
be used to loop over the lattice.

\item[$\bullet$]  Line 07 declares a field of $2\times 2$ matrices, called 
\texttt{phi}, over the lattice \texttt{space}.

\item[$\bullet$]  Lines 08-10 define the matrix A.

\item[$\bullet$]  Lines 11-12 initialize the field \texttt{phi}. Notice that 
\texttt{phi} is distributed over the parallel processes and \texttt{%
forallsites} is a parallel loop.

\item[$\bullet$]  Line 13 performs \emph{synchronization}.

\item[$\bullet$]  Lines 15 through 23 perform 1000 iterations to guarantee
convergence.

\item[$\bullet$]  Line 16 loops over all sites in parallel.

\item[$\bullet$]  Lines 17 through 20 implement eq.~(\ref{eq3}). Notice the
similarity in notation. Here \texttt{phi(x)} is a $2\times 2$ complex matrix.

\item[$\bullet$]  Line 21 performs \emph{synchronization}.

\item[$\bullet$]  Line 23 saves the field. Notice that all fields, including
user-defined ones, inherit \texttt{save} and \texttt{load} methods from the
basic \texttt{mdp\_field} class.
\end{itemize}

\section{Example: Parallel Ising Model}

As one more example of usage of \texttt{MDP} we report here a simple program
for the Ising model.
\begin{verbatim}
00  #include "mdp.h"
01  void main(int argc, char** argv) {
02    mdp.open_wormholes(argc,argv);
03    int L[]={100};
04    mdp_lattice line(1,L);          // declare the lattice
05    mdp_field<int> spin(line);      // declare the spin variables
06    mdp_site x(line);
07    int dE=0, M=L[0], dM=0;         // E for Energy, M for Magnetization
08    float kappa=2.0;                // inverse temperature
09    forallsites(x) spin(x)=+1;      // set initial conditions
10    while(1) {
11      dM=0;
12      for(int parity=EVEN; parity<=ODD; parity++) {
13        forallsitesofparity(x,parity) {
14          dE=2*spin(x)*(spin(x-0)+spin(x+0));   // compute energy variation
15          if(exp(-kappa*dE)>mdp_random.plain()) // Monte Carlo accept-reject
16            { spin(x)*=-1; dM=dM+2*spin(x); }
17        }
18        spin.update(parity);       // communicate
19      }
20      mdp.add(dM);
21      M=M+dM;                      // compute new value for the energy
22      mdp << "magnetization=" << M << endl;
23    }
24    mdp.close_wormholes();
25  }
\end{verbatim}

In this example:

\begin{itemize}
\item  Lines 3-4 declare a 1D lattice of 100 points (\texttt{line}).

\item  Line 5 declares a field of integers (\texttt{spin}) on the lattice.

\item  Line 7 sets the total magnetization M for this initial spin
configuration.

\item  Line 9 sets the initial configuration: all field variables equal to 1.

\item  Line 14 computes the energy variation (\texttt{dE}) of each Markov
Chain Monte Carlo (MCMC)\ step.

\item  Lines 15-16 perform the Monte Carlo accept-reject. If a change is
accepted the spin at site \texttt{x} is flipped and the total magnetization 
\texttt{M} changes (line 16).
\end{itemize}

Note how at each MCMC step, first the code tries to flip the spins at even
locations then, after it updates the lattice sites, it tries to flip the
spins at odd locations (line 13). This guarantees computation results are
independent on parallelization of the lattice line.

Since this even-odd distinction is common in many lattice algorithms, 
\texttt{MDP} stores all even lattice sites and all odd lattice sites close
together. This speeds up loops over one of the two subsets and also speeds
up communication. In fact, in this example, we are able to limit the
synchronization (update) to the site of a given parity (line 18).

\section{Linear Algebra and Other Tools}

\texttt{MDP} includes a Linear Algebra package and other tools. Some of the
most important classes are:

\begin{itemize}
\item  class \texttt{mdp\_real}, that should be used in place of float or
double;

\item  class \texttt{mdp\_complex}, for complex numbers;

\item  class \texttt{mdp\_array}, for vectors and/or multidimensional
tensors;

\item  class \texttt{mdp\_matrix}, for any kind of complex rectangular
matrix;

\item  class \texttt{mdp\_measure}, for error propagation;

\item  class \texttt{mdp\_jackboot}, a container for jackknife and bootstrap
algorithms.
\end{itemize}

The most notable difference between our linear algebra package and other
existing packages is its natural syntax.

For example:
\begin{verbatim}
mdp_matrix A,B;
A=Random.SU(7);
B=exp(A)+inv(A)*hermitian(A)+5;
\end{verbatim}

reads like

$
\begin{tabular}{l}
$A$ and $B$ are matrices \\ 
$A$ is a random $SU(7)$ matrix \\ 
$B=e^A+A^{-1}A^H+5\cdot \mathbf{1}$%
\end{tabular}
$

Note that each matrix can be resized at will and is resized automatically
when a value is assigned.

\texttt{MDP} includes functions for fitting such as the
Levenberger-Marquardt algorithm.

\section{Lattice, Site, and Field}

An \texttt{mdp\_lattice} is the class that describes the space on which
fields are defined; it stores the \emph{topology} of the space (by default
that of a torus in $d$ dimensions) and information about \emph{partioning}
of the space over the parallel processes. In \texttt{MDP}, a lattice is a
graph, defined as a collection of points (lattice \emph{sites}) connected by
links (they specify the topology). Each site is uniquely mapped to one of
the parallel processes.

The only restriction is that the graph must have a degree less than 20. From
now on we will assume the default topology of a torus; therefore the lattice
should be thought of as a mesh in $d\leq 10$ dimensions.

The constructor class \texttt{mdp\_lattice} determines on which process to
store each site, determines the neighbors of each site, and determines the
sizes of the buffers where each process keeps copies of those sites that are
non-local but are neighbors of the local sites.

The constructor also allocates a parallel random number generator so that
each site of the lattice has its own independent random number generator.
This is important for parallel Monte Carlo applications of \texttt{MDP} and
ensures reproducibility of computations on different architectures.

On each lattice it is possible to allocate fields. Some fields are built-in,
for example
\begin{verbatim}
mdp_complex_field
\end{verbatim}

i.e. the field of complex numbers.\ The user can declare any type of field.
For example a field of 5 float per lattice site:
\begin{verbatim}
class S {
 public: float S[5];
};
int L[]={10,10,10};
mdp_lattice cube(3,L);
mdp_field<S> psi(cube);
\end{verbatim}

This code declares a $10\times 10\times 10$ lattice (\texttt{cube}) and a
field (\texttt{psi}), that lives on the \texttt{cube}. The site variables of 
\texttt{psi}, \texttt{psi(x)} belong to class \texttt{S} (assuming \texttt{x}
is an \texttt{mdp\_site} on the \texttt{cube}).

User-defined fields can be saved, loaded, and synchronized
\begin{verbatim}
psi.save("filename");
psi.load("filename");
psi.update();
\end{verbatim}

Synchronization means that all processes will perform MPI communications to
make sure all buffers that contain copies of non-local site variables are
updated with the proper values. The method update should be called
immediately after the local site variables of a field have been changed.

Once a field object is declared in the field constructor, each process
dynamically allocates memory for the buffers that store the copies of those
sites that are non-local, but are neighbors of the local sites. These
buffers are created in such as way to ensure optimal communication patterns.

Every time a field changes, for example in a parallel loop such as
\begin{verbatim}
forallsites(x) phi(x)=0;
\end{verbatim}

the program notifies the field that its values have been changed by calling
\begin{verbatim}
phi.update();
\end{verbatim}

The method \texttt{update} performs all required communication to copy site
variables that need to be synchronized between each couple of overlapping
processes.

Lattice sites are represented by objects of class \texttt{mdp\_site}. Site
objects can be looped over in parallel loops (such as \texttt{forallsites})
but it also possible to explicitly address a specific site by specifying the
site coordinates. Obviously, only the process that stores a site locally
should address a specific site. Class \texttt{mdp\_site} has methods to
check if a site is local, if it is non-local but a local copy is present,
and which process stores the site locally.

\section{Optimal Communication Patterns}

In \texttt{MDP}, the lattice objects, according to the lattice topology and
the parallel partitioning, determine the optimal way to store site variables
in memory and performing parallel communication. This information is then
used by the field method \texttt{update} that performs the synchronization
of the field variables.

Note that \texttt{MDP} does not attempt to overlap computation and
communication. By ``optimal communication'' pattern we mean that, under the
assumptions below, the method update minimizes network traffic and data
copies.

Current optimizations are based on the assumption that each processing node
has one and only one network card and that the network is isotropic (latency
and bandwidth for each couple of nodes is the same). This assumption is
generally true for Ethernet and Myrinet clusters.

Communications are optimal in the sense that:

\begin{itemize}
\item  Each process retrieves all non-local site variables in a single
send/recv for each process that contains sites which are neighbors of local
sites.

\item  Two processes that do not store neighbor sites do not communicate
with each other.

\item  No process is involved in more than a single send and a single
receive at one time.

\item  Each process stores close in memory those copies of non-local site
variables which are local to the same process. In this way synchronization
does not require the use of additional buffers receiving buffers.
\end{itemize}

This technque is particularly efficient for algorithms that only require
next-neighbor communication and run on a all-to-all network topology such as
Ethernet or Myrinet.

It is possible, in principle, to change the above communication patterns to
optimize communication for other network topologies.

Although communications are currently based on MPI, they do not make use of
communication tags. It is therefore possible, in principle, to speed-up
communication by using a faster tagless and bufferless protocol such as
Myricom GM.

Our communication patterns have the effect of making communication almost
insensitive to network latency, and communication speed is dominated by
network bandwidth. Benchmarks are very much application dependent since
parallel efficiency is greatly affected by the lattice size, by the amount
of computation performed per site, processor speed and type of
interconnection. In many typical applications, like the one described in the
preceding example, the drop in efficiency is less than 10\% up to 8 nodes
(processes) and less than 20\% up to 32 (our tests are usually performed on
a cluster of Pentium 4 PCs (2.2GHz) running Linux and connected by Myrinet).

\section{MDP and PSIM}

For portability reasons, \texttt{MDP} is based on MPI. Nevertheless it is
desirable to be able to run, test and debug \texttt{MDP}\ programs on a
single node (with single or multi processor architecture) without having to
install MPI. The latest version of \texttt{MDP} includes a Parallel
SIMulator (PSIM). Despite the name, this is not quite a simulator but an
emulator, i.e., a message passing library that uses local (unix/posix)
socket pairs. PSIM is Objected Oriented and is not based on MPI.

When compiling with PSIM, the parallel processes are created at start-up by
forking. The number of parallel processes is specified at runtime by passing
the following command line argument to any \texttt{MDP} executable program,
\begin{verbatim}
-PSIM_NPROCS=4
\end{verbatim}

(this makes 4 parallel processes).

PSIM also creates a communication log that can be used for debugging. 
\texttt{MDP} with PSIM has been tested on Linux, Mac and Windows (with
cygwin).

For single processor node, using PSIM does not introduce any speed-up, but,
for a small number of processes (2-16) it does not slow down the code
either. For multi-process shared memory architecture, parallelization of
PSIM should produce a speed-up comparable with MPI. We have not been yet
performed such tests.

Moreover, PSIM should perform well on openMosix clusters as soon as
openMosix starts supporting migratable sockets since all communication
between the parallel processes will be done by the operating system.
Unfortunately openMosix does not support migratable sockets yet.

\section{Conclusions}

\texttt{MDP}\ is an easy, powerful, and reliable tool for developing
efficient parallel numerical applications. We have shown here examples in
Computer Science (Cellular Automata), Mathematics (PDE solver) and Physics\
(Ising model).

\texttt{MDP} enables the programmer to focus on algorithm design while
parallelization issues are dealt with automatically in a way transparent to
the programmer.

The underlying communication functions are written in MPI but it is possible
also to compile it without MPI using the built-in PSIM emulator which
enables one to run parallel processes on a single node and/or single
processor architectures, such as a PC. This is useful for testing and
debugging purposes.

All of the features here described are fully functional and have been tested
in real-life applications such as FermiQCD, developed by the University of
Southampton (UK) and Fermilab (Department of Energy).

\smallskip \textbf{\smallskip Acknowledgements}

The author wishes to thank DePaul University for supporting his research and
Fermilab for the fruitful collaboration based on it. He also thanks John
Novak for insightful comments about this manuscript.

\smallskip \textbf{Code download}

\smallskip MDP, PSIM and FermiQCD are currently distributed together and can
be downloaded from:

\texttt{http://www.fermiqcd.net}

\end{document}